\documentclass{ws-procs9x6}

\begin{document}

\title{Einstein constraints on a characteristic cone}

\author{Y. CHOQUET-BRUHAT}
\address{
Acad\'emie des Sciences, Paris
}

\author{P. T. CHRU\'{S}CIEL}
\address{
F\'{e}d\'{e}ration Denis Poisson, LMPT, Tours \\
Hertford College and Oxford Centre for Nonlinear PDE, University of Oxford
}

\author{J. M.  MART\'{I}N-GARC\'{I}A}
\address{
Laboratoire Univers et Th\'eories, CNRS, Meudon, and
Universit\'e Paris Diderot \\
Institut d'Astrophysique de Paris, CNRS, and
Universit\'e Pierre et Marie Curie
}

\begin{abstract}
We analyse the Cauchy problem on a characteristic cone,
including its vertex, for the Einstein equations in
arbitrary dimensions. We use a wave map gauge, solve
the obtained constraints and show gauge conservation.
\end{abstract}


\bodymatter

\section{Introduction}

The Einstein equations in vacuum express the vanishing of the Ricci tensor
of a Lorentzian metric. They are interesting in arbitrary dimensions for the
mathematician and for the physicist looking for unification of the
fundamental interactions. For the analyst they appear as a system of $N$
partial differential equations for the $N$ components of the metric, this
system is invariant by diffeomorphisms, and badly posed, having a
characteristic determinant identically zero. It has been known for a long
time that the Cauchy problem with data on a spacelike hypersurface splits
into constraints to be satisfied by initial data, which can be formulated as
an elliptic system acting on free arbitrary data, and a hyperbolic evolution
system which depends on a gauge choice. An interesting physical problem is
for data on a characteristic cone, i.e. the boundary of the future of one
point, generated by the light rays issued from that point. The problem has
common and also different aspects from the spacelike Cauchy problem.
Compared with the case of initial data prescribed on a pair of intersecting
null hypersurfaces treated before in spacetime dimension four \cite{AR,DS}
it presents new mathematical difficulties due to its singularity at the
vertex, and only partial results \cite{CB1} have been obtained before, with no
treatment of the light-cone constraints so far, even in dimension four. We
present an approach to the solution of the Einstein equations with data on a
characteristic cone in all dimensions $n+1\geq 3$ using a wave-map gauge.

\section{Wave-map gauge.}

Harmonic coordinates have been used for a long time to study evolution of
solutions of the Einstein equations, they are only locally defined in
general, and non tensorial. The wave-map gauge provides conditions which are
tensorial and global in space. A metric $g$ on a manifold $V$ will be said
to be in $\hat{g}$\textbf{-wave-map gauge} if the identity map $%
V\rightarrow V$ is a harmonic diffeomorphism from the spacetime $(V,g)$ onto
the pseudo-Riemannian manifold ($V,\hat{g})$ (see Ref. \refcite{CB2}). The \textbf{%
wave-gauge vector} $H$ is given in arbitrary coordinates by the formula,
with $\Gamma _{\alpha \beta }^{\lambda }$ and $\hat{\Gamma}_{\alpha \beta
}^{\lambda }$ the Christoffel symbols of the metrics $g$ and $\hat{g},$
\begin{equation}
H^{\lambda }:=\Gamma ^{\lambda }-W^{\lambda }\;,\ \text{with}\ \ \ \ \Gamma
^{\lambda }:=g^{\alpha \beta }\Gamma _{\alpha \beta }^{\lambda },\text{ \ \ }%
W^{\lambda }:=g^{\alpha \beta }\hat{\Gamma}_{\alpha \beta }^{\lambda }.
\end{equation}
The Ricci tensor of the spacetime metric $g$ satisfies the identity with $%
\hat{D}$ the covariant derivative in the metric $\hat{g}$~
\begin{equation}
R_{\alpha \beta }\equiv R_{\alpha \beta }^{(H)}+{\frac{1}{2}}(g_{\alpha
\lambda }\hat{D}_{\beta }H^{\lambda }+g_{\beta \lambda }\hat{D}_{\alpha
}H^{\lambda })  \label{RicciHIdentity}
\end{equation}
where $R_{\alpha \beta }^{(H)}(g)$, called the reduced Ricci tensor, is a
quasi-linear, quasi-diagonal operator on $g$, tensor-valued, depending on $%
\hat{g}$.

In the case in which the target metric is the Minkowski metric $\eta $, with
covariant derivative denoted by $D$, the reduced Einstein equations read in
vacuum
\begin{equation}  \label{redEin}
Ricci^{(H)}(g)\equiv -{\frac{1}{2}}g^{\lambda \mu }D_{\lambda }D_{\mu
}g+Q(g)(Dg,Dg)=0\;,
\end{equation}
with $Q(g)(Dg,Dg)$ a quadratic form in $Dg$ with coefficients analytic in $g$
as long as $g$ is not degenerate.

\section{Characteristic Cauchy problem for quasilinear wave equations.}

The reduced Einstein equations  {(\ref{redEin})} form a
quasi-linear, quasi-diagonal system of wave equations for the
Lorentzian metric $g$. It is known since Leray's work \cite{JL} on
hyperbolic systems, that, in the linear case, the Cauchy
problem for such systems on a given globally hyperbolic
spacetime is well posed if the initial manifold $M_{0}$ is
``compact towards the past''; that is, is intersected along a
compact set by the past of any compact subset of $V$. However
the data depends on the nature of $M_{0}$ and the formulation
of a theorem requires care. In the case of a wave equation with
$M_{0}$ a null hypersurface, except at some singular subsets
(intersection in the case of two null hypersurfaces, vertex in
the case of a null cone) the data is only the function, not its
transversal derivative,
with some hypotheses which need to be made as one approaches the singular {%
set}.

Following a theorem proved by Cagnac \cite{FC}, Dossa \cite{MD} proved well posedness
of the Cauchy problem for quasilinear wave equations of the form
\begin{equation}
A^{\lambda \mu }(y,v)\partial _{\lambda \mu }^{2}v+f(y,v,\partial v)=0,\quad
y=(y^{\lambda })\in \mathbf{R}^{n+1},\quad \partial _{\lambda }:=\frac{%
\partial }{\partial y^{\lambda }},  \label{Dossa1}
\end{equation}%
with initial data given on a subset, including its vertex $O$, of a
characteristic cone $C_{O}$
\begin{equation}
\bar{v}:=v|_{C_{O}}=\phi ,  \label{3.2}
\end{equation}%
which is the restriction of a smooth spacetime function.

Both Cagnac and Dossa assume that there is a domain $U\subset \mathbf{R}%
^{n+1}$ where the characteristic cone $C_{O}$ of vertex $O$ is represented
by the cone\footnote{{\footnotesize It is known, using normal coordinates,
that a null cone in a $C^{1,1}$ Lorentzian spacetime always admits such a
representation in a neighbourhood of its vertex; the null geodesics which
generate it are represented by the lines where $r$ only varies.}} in $%
\mathbf{R}^{n+1}$
\begin{equation*}
C_{O}:=\{x^{0}\equiv r-y^{0}=0\}\;,\quad r^{2}:=\sum_{i=1,\ldots
,n}(y^{i})^{2}.
\end{equation*}%
The initial data $\phi $ is assumed to be defined on the domain
\begin{equation*}
C_{O}^{T}:=C_{O}\cap \{0\leq t:=y^{0}\leq T\}.
\end{equation*}%
One denotes by $Y_{O}:=\{t:=y^{0}>r\}$ the interior of\ $C_{O}\;$\ and sets
\begin{align}
& \text{ \ }Y_{O}^{T}:=Y_{O}\cap \{0\leq y^{0}\leq T\}\;, \\
& \Sigma _{t}:=C_{O}\cap \{y^{0}=t\}\;,\quad \text{diffeomorphic to\ }%
S^{n-1}\;, \\
S_{t}& :=Y_{O}\cap \{y^{0}=t\}\;,\quad \text{diffeomorphic to the ball\ }%
B^{n-1}\;.
\end{align}%
An existence theorem which can be used for the Einstein equations is

\begin{theorem}
 \label{TCD}
(Cagnac-Dossa) There is a number $T_{0}>0$ such that the
problem (\ref{Dossa1})-(\ref{3.2}) has one and only one solution $v$,
smooth in $Y_{O}^{T_{0}},$ if

1. There is an open set $U\times W\subset \mathbf{R}^{n+1}\times \mathbf{R}%
^{N}$, $Y_{O}^{T}\subset U$ where the functions $A^{\lambda \mu }$ are
smooth in $y$ and $v$, the quadratic form $A^{\lambda \mu }$ has Lorentzian
signature; it takes the Minkowskian values for $v=0$.

2. a. The function $\phi $ takes its values in $W.$ The cone $C_{O}^{T}$ is
null for the metric $A^{\lambda \mu }(y,\phi )$ and $\phi (O)=0$.

b. $\phi $ is the trace on $C_{O}^{T}$ of a smooth function in $U.$
\end{theorem}

More refined hypotheses and conclusions on functional spaces for data and
solutions are given in Ref. \refcite{MD}, but there seems to be some difficulty to apply
the result to initial data admissible for the Einstein equations.

\section{The case of Einstein equations.}

{Theorem~\ref{TCD}} applies to the reduced Einstein equations
in vacuum {(\ref{redEin})} on $\mathbf{R}^{n+1}$: these
equations take the form {(\ref{Dossa1})} for the unknown
$v=g-\eta $ if the target of the wave-map is the Minkowski
spacetime whose metric $\eta $ takes in the coordinates $y$ of
$\mathbf{R}^{n+1}$ the canonical form
\begin{equation}  \label{Mink}
\eta \equiv -(dy^{0})^{2}+\sum (dy^{i})^{2}.\text{ }
\end{equation}

We formulate the theorem as follows.

\begin{theorem}
Let $\bar{g}$ be the trace on $C_{O}^{T}$ of a Lorentzian metric $%
G$ smooth in a neighbourhood of $C_{O}^{T}$ and such that $G(O)=\eta .$ Then
there exists $T_{0},$ $0<T_{0}\leq T,$ such that the reduced vacuum Einstein
equations $Ricci(g)^{(H)}=0$ admit one and only one solution in $%
Y_{O}^{T_{0}},$ which is a smooth Lorentzian metric $g^{(H)}=\eta +h$, with $%
h$ taking the value $\bar{G}-\bar{\eta}$ on $C_{O}^{T_{0}}.$
\end{theorem}

The problem is now to find conditions under which a solution of the reduced
equations satisfy the full Einstein equations.

\begin{theorem}
\label{Existence2} A Lorentzian metric $g$ solution of the reduced Einstein
equations in wave-map gauge $Ricci^{(H)}(g)=0$ in $Y_{O}^{T_{0}}$ is a
solution of the full Einstein equations $Ricci(g)=0$ if the wave gauge
vector $H$ vanishes on $C_{O}^{T_{0}}$.
\end{theorem}

\noindent

\begin{proof}
The Bianchi identities
\begin{equation*}
\nabla _{\alpha }S^{\alpha \beta }\equiv 0,\text{ \ \ \ }S^{\alpha \beta }:=%
{R}^{\alpha \beta }-\frac{1}{2}g^{\alpha \beta }R
\end{equation*}
imply that for a smooth solution of the reduced Einstein equations the
wave-gauge vector $H$ satisfies the quasidiagonal linear homogeneous system
of second order equations with principal term the wave operator in the
metric $g$,
\begin{equation}
\nabla _{\alpha }D^{\alpha }H^{\beta }+\nabla _{\alpha }D^{\beta }H^{\alpha
}-\nabla ^{\beta }D_{\alpha }H^{\alpha }=0.  \label{WaveH}
\end{equation}
These equations imply that $H=0$ in $Y_{O}^{T_{0}}$ if its trace $\bar{H}$
on $C_{O}^{T_{0}}$ vanishes.
\end{proof}

\medskip

One has to find conditions under which the trace $\bar{H}$ on $C_{O}^{T_{0}}$
of the wave gauge vector $H$ vanishes: these will lead to constraints on the
initial data.

\section{ Adapted null coordinates.}

Two facts have to be stressed

1. The values $\bar{H}$ on $C_{O}$ of the wave gauge vector $H$ depend not
only on the initial data $\bar{g},$ but also on derivatives transversal to
the cone.

2. The trace $\bar{g}$ of a spacetime metric $g$ is not invariant under
isometries preserving $C_{O}.$

The following theorem has some analogies with the theorem
concerning the case of spacelike initial data.

We denote by $\bar{S}_{\alpha \beta }:=\bar{R}_{\alpha \beta }-\frac{1}{2}%
\bar{g}_{\alpha \beta }\bar{R}$ the trace on $C_{O}$ of the Einstein tensor,
by $\ell $ a field of null vector fields tangent (and also normal) to $C_{O}$%
.

\begin{theorem}
\label{Existence3} The operator $\bar{S}_{\alpha \beta }\ell ^{\beta }$ is a
sum $\mathcal{L}_{\alpha }+\mathcal{C}_{\alpha }$\ where $\mathcal{L}%
_{\alpha }$ vanishes when $\bar{H}=0$ while $\mathcal{C}_{\alpha }$ depends
only on the data $\bar{g}$ and the choice of the target space of the gauge
wave map.
\end{theorem}

An immediate consequence of this theorem is that a Lorentzian metric $g$
taking an initial value $\bar{g}$ on a null cone $C_{O}^{T}$ and solution of
the reduced Einstein equations in wave gauge in the future of $C_{O}^{T}$ is
solution of the original Einstein equations only if $\bar{g}$ satisfies the
constraints $\mathcal{C}_{\alpha }=0$. The converse statement will be a
consequence of the explicit expressions of the $\mathcal{L}_{\alpha }.$

The proof of Theorem~\ref{Existence3} requires the introduction
of adapted coordinates.

\bigskip

\textbf{Coordinates adapted to a null cone.}

The $y^{\alpha }$ are coordinates defining the differentiable structure of $%
\mathbf{R}^{n+1},$ : $t\equiv y^{0},y^{i},$ $i=1,...n,$
$r^{2}:=\Sigma _{i}(y^{i})^{2},$ we have chosen the target
Minkowski metric to be given by {(\ref{Mink})} in these
coordinates.

The $x^{\alpha }$ coordinates, adapted to the null structure of the cone $%
C_{O}$: $x^{0}=r-y^{0},$ are defined to be $x^{1}=r,$ $x^{A}$ local
coordinates on the sphere $S^{n-1}.$ On the null cone $C_{O}:=\{x^{0}=0\}$
of a Lorentzian metric $g$ the trace $\bar{g}$ of $g$ on $C_{O}$ is such
that $\bar{g}^{00}=0$ . If the coordinates $x^{1}$ and $x^{A}$ are chosen
such that on $C_{O}$ it holds that $\frac{\partial }{\partial x^{1}}$ $%
\equiv \ell ,$ the null direction, then $\bar{g}_{11}=0$ and $\bar{g}%
_{1A}=0, $ hence $\bar{g}^{0A}=0;$ that is, the trace $\bar{g}$ reduces to
\begin{equation*}
\bar{g}:=\bar{g}_{00}(dx^{0})^{2}+2\nu _{0}dx^{0}dx^{1}+2\nu
_{A}dx^{A}dx^{1}+\tilde{g},
\end{equation*}
where \thinspace $\tilde{g}:=\bar{g}_{AB}dx^{A}dx^{B}$ is a positive
definite quadratic form induced on $C_{O}$ by the embedding of $g$ in $(%
\mathbf{R}^{n+1},g).$ In the $x$  {coordinates} the Minkowski
metric  {(\ref{Mink})} reads
\begin{equation*}
\eta :=-(dx^{0})^{2}+2dx^{0}dx^{1}+(x^{1})^{2}s_{n-1},
\end{equation*}
with $s_{n-1}:=s_{AB}dx^{A}dx^{B}$ the metric of the unit sphere $S^{n-1}.$

Remark that while $\tilde{g}$ is invariant under isometries of $g,$ the
other components of $\bar{g}$ are gauge dependent.

The following fundamental theorem completes Theorem~{\ref{Existence3}%
}. It holds on any null hypersurface\footnote{{\footnotesize For dimension }$%
n=3${\footnotesize \ and harmonic coordinates the part 1 is proved in Ref. \refcite{AR}
and the constraints }$C_{\alpha }${\footnotesize \ explicitly constructed in
 Ref. \refcite{DS}.}}.

\begin{theorem}
In adapted null coordinates:

1. The operators $\mathcal{C}_{\alpha }$ are a hierarchy of ordinary
differential operators for the coefficients of $\bar{g}$ along the
generators.

2. The operators $\mathcal{L}_{\alpha }$  {lead to}  a
hierarchy of linear homogeneous first order operators for the
components $\bar{H}_{\alpha }$ if the spacetime metric $g$
satisfies on $C_{O}$ the reduced Einstein equations.
\end{theorem}

To write the $\mathcal{C}_{\alpha }$ and $\mathcal{L}_{\alpha }$ one
introduces the null second fundamental form, Lie derivative of $\tilde{g}$
with respect to $\ell ,$ given in adapted coordinates by
\begin{equation}
\chi _{AB}:=\overline{\nabla _{A}\ell _{B}}\equiv \frac{1}{2}\partial _{1}%
\bar{g}_{AB}.
\end{equation}%
and denotes by $\tau $ the mean null extrinsic curvature
\begin{equation}
\tau :=\frac{1}{2}\bar{g}^{AB}\partial _{1}\bar{g}_{AB}.
\end{equation}

\section{$\mathcal{C}_{1}$ constraint.}

The first constraint $\mathcal{C}_{1}=0$ is

\begin{equation*}
\bar{S}_{\alpha \beta }\ell ^{\alpha }\ell ^{\beta }\equiv \bar{S}%
_{11}\equiv \bar{R}_{11}=0.
\end{equation*}
Computation gives (indices raised with $\tilde{g}),$ \ with $\bar{\Gamma}%
_{1}\equiv $ $\bar{W}_{1}+\bar{H}_{1}$%
\begin{equation*}
\bar{R}_{11}\equiv -\partial _{1}\tau +\nu ^{0}\partial _{1}\nu _{0}\tau -%
\frac{1}{2}\tau (\bar{\Gamma}_{1}+\tau )-\chi _{A}{}^{B}\chi _{B}{}^{A}
\end{equation*}
hence
\begin{equation*}
\bar{R}_{11}\equiv \mathcal{C}_{1}+\mathcal{L}_{1},\text{ \ \ \ \ }\mathcal{L%
}_{1}\equiv -\frac{1}{2}\tau \bar{H}_{1},\text{\ }
\end{equation*}
and denoting by $\sigma $ denotes the traceless part of $\chi ,$

\begin{equation*}
\sigma _{A}^{B}:=\chi _{A}^{B}-\frac{1}{n-1}\delta _{A}^{B}\tau \text{ \ \ \
\ \ \ }|\sigma |^{2}:=\sigma _{A}^{B}\sigma _{B}^{A},\text{\ }
\end{equation*}
\begin{equation*}
\mathcal{C}_{1}\equiv -\partial _{1}\tau -\frac{\tau ^{2}}{n-1}+\tau \{\nu
^{0}\partial _{1}\nu _{0}-\frac{1}{2}(\bar{W}_{1}+\tau )\}-|\sigma |^{2}=0,
\end{equation*}
We remark that $\sigma _{A}^{B}$ depends only on the conformal class of $%
\tilde{g}$; if $\tilde{g}\equiv \varphi ^{2}\gamma $ then
\begin{equation*}
\sigma _{A}^{B}\equiv \frac{1}{2}\{\gamma ^{BC}\partial _{1}\gamma _{AC}-%
\frac{1}{n-1}\delta _{A}^{B}\partial _{1}\log |\gamma |\},\text{ \ \ }%
|\gamma |:=\det \gamma
\end{equation*}
The free data will be a representant $\gamma $ of this conformal class.

\subsection{Solution of the $\mathcal{C}_{1}$ constraint.}

One can replace the constraint $\mathcal{C}_{1}$ by the two
equations, one containing only $\tau $
\begin{equation}
\partial _{1}\tau +\frac{\tau ^{2}}{n-1}+|\sigma |^{2}=0
,
\end{equation}%
and an equation for $\nu _{0}$  which can be solved only after $\tilde{g}$%
, i.e. a conformal factor $\varphi $ deduced from $\tau,$ is determined%
\begin{equation}
\nu ^{0}\partial _{1}\nu _{0}-\frac{1}{2}(\bar{W}_{1}+\tau )=0,\text{ \ \
with \ \ }\nu _{0}(O)=1
.
 \label{nueq}
\end{equation}%
When $|\sigma |^{2}=0$ the solution corresponding to a Minkowski cone $C_{O}$
is $\tau =\frac{n-1}{r}.$ In the general case one proves existence on $%
C_{O}^{T}$ of a positive $\tau ,$ with $\frac{n-1}{r}-\tau $ tending to zero
at $O,$ if $T$ is small enough; $T=\infty $ if $%
\int_{0}^{\infty }r^{2}|\sigma |^{2}dr$ is small enough. One also
sees that if there exists $r_{1}>0$ such that
\begin{equation*}
\int_{0}^{r_{1}}r^{2}|\sigma |^{2}(r,x^{A})d\rho \geq (n-1)r_{1}\;,
\end{equation*}
then $\tau $ vanishes at a finite $r\geq r_{1}$ along this
generator, possibly leading then to an outer trapped surface.

Knowing $\tau $ one uses a first order equation for $\varphi $,
\begin{equation*}
\partial _{1}\log \varphi =\frac{1}{n-1}(\tau -\partial _{1}(\log \det
\gamma )^{\frac{1}{2}}),\text{ \ \ with \ }\varphi (0)=1.
\end{equation*}

Knowing $\varphi ,$ hence $\tilde{g}:=\varphi ^{2}\gamma ,$ one
shows that (\ref{nueq}) has one and only one smooth
solution tending to 1 at
the vertex $O.$ This $\nu _{0}$ has no zeros on $C_{O}$ if $%
\int_{0}^{\infty }r^{2}|\sigma |^{2}dr$ is small enough.

\subsection{\textbf{Vanishing of }$\bar{H}_{1}.$}

The identities
\begin{equation*}
\bar{R}_{11}\equiv \bar{R}_{11}^{(H)}+\nu _{0}D_{1}\bar{H}^{0}\equiv
\mathcal{C}_{1}+\mathcal{L}_{1},\text{ \ \ }\bar{H}_{1}\equiv \nu _{0}\bar{H}%
^{0}
\end{equation*}%
show that a solution of the characteristic Cauchy problem with data on $C_{O}
$ for the reduced Einstein equations, i.e. such that $\bar{R}_{11}^{(H)}=0$,
and initial data satisfying the constraint $\mathcal{C}_{1}$ is such that $%
\bar{H}^{0}$ satisfies a linear homogeneous differential equation on $C_{O}$%
, which implies $\bar{H}^{0}=0$,
\begin{equation*}
D_{1}\bar{H}^{0}+\frac{1}{2}\bar{H}^{0}\tau =0.
\end{equation*}

\noindent\textbf{Remark:} Since $\bar{H}_{1}\equiv \nu
_{0}\bar{H}^{0}=0$ the equation chosen for $\nu _{0}$ implies
that $x^{1}$ is an affine parameter on the null geodesic. The
$\mathcal{C}_{1}$ constraint appears then as the gauge
independent Raychaudhuri equation.

\section{$C_{A}$\ and $C_{0}$ constraints.}

One finds
\begin{equation*}
\bar{S}_{1A}\equiv \bar{R}_{1A}\equiv \mathcal{C}_{A}+\mathcal{L}_{A},
\end{equation*}%
with, when $\bar{H}_{1}=0,$%
\begin{equation*}
\mathcal{L}_{A}\equiv -\frac{1}{2}\{\partial _{1}\bar{H}_{A}+\tau \bar{H}%
_{A}\}
\end{equation*}%
while $\mathcal{C}_{A}$ is a linear second order differential operator%
\footnote{{\footnotesize The explicit formulas have been obtained by
algebraic computing \cite{MG}.}} along the generators $\ell $ with only
unknown $\nu _{A}$ when $\tilde{g}$ is known.

Also
\begin{equation*}
\bar{S}_{01}\equiv \mathcal{C}_{0}+\mathcal{L}_{0},
\end{equation*}
with when $\mathcal{C}_{1}=\mathcal{C}_{A}=\mathcal{L}_{1}=\mathcal{L}_{A}=0$

\begin{equation*}
\mathcal{L}_{0}\equiv -\partial _{1}\bar{H}_{0}+\frac{1}{2}(\bar{W}%
_{1}-\tau )\bar{H}_{0}
\end{equation*}
and $\mathcal{C}_{0}$ a linear second order differential operator along the
generators $\ell $ for the only remaining unknown $\bar{g}_{00}.$

The expressions of the $\mathcal{L}$'s together with the identities
(\ref{WaveH}) allow to prove the vanishing of the $\bar{H}$'s. The
differential equations for $\nu _{A}$ and $\alpha
:=\bar{g}_{00}+1$ have singularities at the vertex $x^{1}=0$
but can be proved to have solutions vanishing at the
vertex together by the product by $x^{1}$ of their first
derivatives. These solutions are global and give a Lorentzian
metric  as long as $\tau
>0$ and $\nu _{0}>0$.

\section{\textbf{Conclusions. }}

\subsection{Existence.}

The Einstein wave map gauge constraints have a solution on
$C_{O}^{T_{0}}$ if $T_{0}$ is small enough. The solution is
global on $C_{O}$ if $ \int_{0}^{\infty }r^{2}|\sigma |^{2}dr$
is small enough.

In the vacuum case the problem of evolution has then a smooth solution in
$Y_{O}^{T_{0}}$ if the constructed initial data $\bar{g}$ is the trace on
$C_{O}$ of a smooth Lorentzian metric. We conjecture that this
property holds when the free data $\gamma $ is induced on
$C_{O}$ by a smooth Lorentzian metric.
An explicit necessary and sufficient condition  on
$\gamma$ for a $C^2$ solution can be given in terms of tensor
spherical harmonics.

\subsection{Local geometric uniqueness.}

The components of the trace $\bar{g}$ on a codimension $1$ submanifold $M$
of a metric $g$ of a manifold $V$ are defined in arbitrary coordinates, but
they don't obey a tensorial transformation law under a diffeomorphism which
preserves $M.$ An invariant geometric quantity on $M$ is the pull back of $g$
by the identity map $f$\ given by the embedding of $M$ into $V;$ that is,
the quadratic form $\tilde{g}$ in the case where $M$ is the null submanifold
$C_{0}.$

The previous results lead to the following theorem.

\begin{theorem}
A spacetime $(V,g)$ solution of the vacuum Einstein equations in the future
of a characteristic cone is uniquely determined, up to isometry, by the
conformal class of the degenerate quadratic form induced on $C_{O}$ by $g.$
\end{theorem}

\section{\textbf{Open problems}}

Analytic rigorous study of the asymptotic behaviour as $r$ tends to infinity.

Global existence of evolution for small data.

Formation of trapped surfaces and black holes (problem treated in Ref. \refcite{DC} for $%
n=3$ and data Minkowskian in a neighbourhood of the vertex).

\bibliographystyle{ws-procs9x6}

\end{document}